\documentclass[english,aps,prb,showpacs,twocolumn]{revtex4}
\usepackage{graphicx}
\usepackage{babel}

\def\ket#1{\left|#1 \right\rangle}
\def\meanval#1{{\left\langle #1 \right\rangle}}
\def\braket#1#2{\left\langle #1 | #2 \right\rangle}
\def\meanvalbraket#1#2#3{\left\langle #1 | #2 | #3 \right\rangle}
\def\mod#1{{\left|#1\right|}}
\def\disor#1{\left\langle #1 \right\rangle_{\rm aver}}
\begin{document}

\title{Broadening effects due to alloy scattering in Quantum Cascade Lasers}

\author{N.~Regnault}

\author{R.~Ferreira}

\author{G.~Bastard}

\affiliation{Laboratoire Pierre Aigrain - Ecole Normale Sup\'erieure,
24 rue Lhomond, F-75005 Paris, France}


\begin{abstract}

We report on calculations of broadening effects in QCL due to alloy scattering.  The output of numerical calculations of alloy broadened Landau levels compare favorably with calculations performed at the self-consistent Born approximation.  Results for Landau level width and optical absorption are presented.  A disorder activated forbidden transition becomes significant in the vicinity of crossings of Landau levels which belong to different subbands.  A study of the time dependent survival probability in the lowest Landau level of the excited subband is performed.  It is shown that at resonance the population relaxation occurs in a subpicosecond scale.

\end{abstract}


\maketitle

\section{Introduction}

A salient feature of the Quantum Cascade Lasers (QCL) is the modulation of their light output by a magnetic field applied parallel to the growth axis\cite{Becker02,Alton03,Tamosiunas03,Ulrich00,Blaser02}.  This modulation can even be so strong that the laser turns off, a phenomenon refereed to as QCL blinking.  The magnetic field Landau quantizes the electron in - plane motion and the laser output shows a resonant decrease whenever the ground Landau level of the upper state of the lasing transition becomes energy lined up with an excited LL of the lower (or intermediate) state or with one LO phonon replica of such an excited LL.
The scattering effects between LL's of a 2D gas are singular because of the unperturbed LL's are evenly spaced and macroscopically degenerate.  They have been mostly studied inside the ground subband for degenerate electrons (the situation of the Quantum Hall Effect, see e.g. \cite{Yoshioka}).  The blinking QCL's are particularly interesting because the scatterings produce spectacular effects and involve a change in the subband index and of a LL index for a non degenerate electron population.
In this paper, we report on the calculations of the inter LL, intersubband scattering effects due to alloy fluctuations.  The alloy scattering is known to be a very efficient scattering mechanism in quantum wells \cite{Bastard83} and it is of importance to ascertain its effect for mid infrared QCL's.  Both a numerical study and a semi - analytical study will be performed.  In section II, we discuss our model as well as the numerical tools we have used in the numerical calculations. Section III will be devoted to results for the density of states and the LL widths while in section IV we shall present results on the optical absorption associated with intersubband transitions.  In section V, we discuss the time dependent survival probability of an electron in the ground LL of the upper state.

\section{Model}

We work in the effective mass approximation and want to solve the Schr\"odinger equation

\begin{eqnarray}
{\cal H}\Psi\left(\vec{r},t\right)&=&i\hbar\frac{\partial}{\partial t}\Psi\left(\vec{r},t\right)\label{schroedinger}
\end{eqnarray}
with 
\begin{eqnarray}
{\cal H}&=&\frac{1}{2m^*}\left(p_x^2+\left(p_y+eBx\right)^2+p_z^2\right)\label{hamiltonian}\\
&&+V_{\rm conf}(z)+V_{\rm alloy}(\vec{r})\nonumber
\end{eqnarray}

where we have used the transverse gauge for the vector potential and neglected spin effects.  The $z$ dependent problem admits several subbands\cite{Bastard83} :
\begin{eqnarray}
\left(\frac{p_z^2}{2m^*}+V_{{\rm conf}}(z)\right)\chi_l(z)&=&E_l \chi_l(z)\label{subbanddef}
\end{eqnarray}

For mid infrared QCL's, the subband energy distance is $\simeq 0.1 {\rm eV}$.  Thus, it is a good approximation to assume that the alloy scattering is going to destroy neither the subband structure nor at high field the LL structure. Thus we can consider a basis that includes a few subbands and LL's.  Moreover, we shall replace in the following the elaborate layer sequence of an actual QCL by a single quantum well with infinite barriers.  The well thickness is $L_z$.
The numerical computation is done by diagonalizing the Hamiltonian in a large box ($100{\rm nm}\times 100{\rm nm}\times L_z$). To model the ${\rm Ga}_{1-x}{\rm In}_x {\rm As}$ alloy, we partition the large box into tiny cubes ($5 \AA$ side). In each cube the scattering potential is a random variable equal to $x \Delta V$ with probability $1-x$ and to $-(1-x)\Delta V$ with probability $x$ where $\Delta V = 0.6 {\rm eV}$\cite{Leuliet06} :
\begin{eqnarray}
V_{{\rm alloy}} &=& \sum_{\vec{R}_{\rm Ga}} x \Delta V f\left(\vec{r}-\vec{R}_{\rm Ga}\right)\nonumber\\
&&- \sum_{\vec{R}_{\rm In}} (1-x) \Delta V f\left(\vec{r}-\vec{R}_{\rm In}\right)\label{Valloy}
\end{eqnarray}

where $f(\vec{R}_i)$ is a step function which is one in the tiny cube centered at a lattice site $\vec{R}_i$ and zero elsewhere. On the scale of the Landau levels or the $z$ dependent subbands $\chi_n$ the function $f$ is well approximated by $\omega_0 \delta(\vec{r}-\vec{R}_i)$ where $\omega_0$ is the volume of the unit cell. There is no correlation between the $x$ values of different cubes.  The computer generates a random occupation of the tiny cubes. An average over $N$ such trials is done at the end of the calculations. $N$ was taken equal to 100. A numerical diagonalization is used to extract the eigenvalues and eigenfunctions of the perturbed LL's.  The calculations are done with $m^* = 0.05 m_0$, $x = 0.53$, $E_2 - E_1 = 143 {\rm meV}$ and $L_z=125\AA$. In the following we concentrate on the vicinity of the critical field $B_r$ that lines up the $(2,0)$ LL and the $(1,2)$ LL (we label $(l,n)$ the $n^{\rm th}$ LL related to the $l^{\rm th}$ subband).  With the material parameters we use, this field is $B_c = 30.9{\rm T}$.

\section{Density of states and width of broadened Landau levels}

On the scale of the cyclotron radius or well width, the alloy fluctuations act like delta scatterers. With the Landau gauge, the unperturbed LL eigenstates of the $l$-th $z$ dependent subband are given by:

\begin{eqnarray}
\varphi_{l,n,k_y}\left(\vec{r}\right)&=&\frac{e^{i k_y y}}{\sqrt{L_y}}\chi_l(z)H_n(x+\lambda^2 k_y)\nonumber\\
\varepsilon_{l,n}&=&E_l+\left(n+\frac{1}{2}\right)\hbar\omega_c\label{wavefunction}\\
\lambda^2=\frac{\hbar}{eB}&;&\omega_c=\frac{eB}{m^*}\nonumber
\end{eqnarray}

where $H_n$ is the $n$-th Hermite function. Each of the $(l,n)$ LL carries a degeneracy $\frac{L_x L_y}{2\pi\lambda^2}$ and has an unperturbed energy $\varepsilon_{l,n}$. In our confining box this means $48$ states at $20{\rm T}$.
In the presence of short range scatterers and provided one restrict the analysis to intra-$n$ (and a fortiori intra-$l$) transitions, one gets at the self consistent Born approximation (SCBA; see e.g. \cite{Ando82}) the averaged Green function and self energy : 

\begin{eqnarray}
\meanval{G(\varepsilon)}_{l,n}&=&\frac{1}{\varepsilon_+-\varepsilon_{l,n}-\Sigma_{l,n}(\varepsilon)+i0}\nonumber\\
\Sigma_{l,n}(\varepsilon)&=&\overline{V}^2_{ln,ln}\meanval{G(\varepsilon)}_{l,n} \label{defgreen}\\
V^{k_y,{k'}_y}_{ln,l'n'}&=&\int d^3r \varphi^*_{l,n,k_y}(\vec{r})V_{{\rm alloy}}(\vec{r})\varphi_{l',n',{k'}_y}(\vec{r})\nonumber\\
\overline{V}^2_{ln,ln}&=&\disor{\sum_{k_y'}\mod{V^{k_y,{k'}_y}_{ln,ln}}^2}\nonumber
\end{eqnarray}

where $\disor{\;}$ stands for the average over the alloy fluctuations.  Hence :

\begin{eqnarray}
\Sigma_{l,n}(\varepsilon)&=&\frac{\varepsilon-\varepsilon_{l,n}}{2}-i\sqrt{\overline{V}^2_{ln,ln}-\left(\frac{\varepsilon-\varepsilon_{l,n}}{2}\right)^2}\nonumber
\end{eqnarray}

The averaged density of states (DOS) of the $(l,n)$ broadened LL retains the familiar semi-elliptical shape\cite{Ando82} :

\begin{eqnarray}
\rho_{l,n}\left(\varepsilon\right)&=&-\frac{eBL_xL_y}{2\pi^2\hbar \overline{V}^2_{ln,ln}}{\rm Im}\left(\Sigma_{l,n}(\varepsilon)\right)
\end{eqnarray}

and extends over $\overline{V}_{ln,ln}$ on each side of the unperturbed $(l,n)$ LL energy.  The matrix element $\overline{V}_{ln,ln}$ averaged over the position of the alloy fluctuation is such that :

\begin{eqnarray}
\overline{V}^2_{ln,ln}&=&\frac{x(1-x)\omega_0}{2\pi \lambda^2}\left(\int_{\omega_0} d^3r\;\left|f\left(\vec{r}\right)\right|^2\Delta V\right)^2\nonumber\\
&&\times \int_{-\infty}^{+\infty}dz\;\chi_l^4(z)\\
&=&\frac{x(1-x)\omega_0 \left(\Delta V\right)^2}{2\pi \lambda^2}\int_{-\infty}^{+\infty}dz\;\chi_l^4(z)\nonumber
\end{eqnarray}

As it is well known $V_{ln,ln}$ does not depend on $n$. Note that in our case of an infinitely deep quantum well, $\overline{V}_{ln,ln}$ does not depend on $l$ either since :

\begin{eqnarray}
\int_{-\infty}^{+\infty}dz\;\chi_l^4(z)=\frac{3}{2L_z}&&{\rm for\;all\;}l
\end{eqnarray}

The assumption of this model, namely that the impurity potential has only intra-n,intra-l matrix elements is reasonable in strong magnetic field for the usual 2D gases.  In the case of QCL's, where one is interested in the fate of electrons staying on the $n=0$ LL of an excited subband (called here $E_2$ for the sake of definiteness), this approximation fails when this $n=0$ LL is intersecting one LL (here $n=2$) that belongs to a different subband (called here $E_1$ for the sake of definiteness).

To analyze this situation, one can restrict the Hilbert space to these two close-by LL's and write the wave function of a state:
\begin{eqnarray}
\Psi_{\nu}\left(\vec{r}\right)&=&\sum_{k_y}c_{\nu}(k_y)\varphi_{1,2,k_y}+d_{\nu}(k_y)\varphi_{2,0,k_y}\label{DefPsiNu}
\end{eqnarray}

where $c_{\nu}(k_y)$ and $d_{\nu}(k_y)$ are constant coefficients and $\nu$ labels the eigenstates. If we restrict ourselves to the SCBA we have now to solve :

\begin{eqnarray}
\Sigma_{2,0}(\varepsilon)&=&\disor{\sum_{{k'}_y}V^{k_y {k'}_y}_{20,20}\frac{1}{\varepsilon-\varepsilon_{2,0}-\Sigma_{2,0}(\varepsilon)+i0}V^{{k'}_y k_y}_{20,20}}\nonumber\\
&+&\disor{\sum_{{k'}_y}V^{k_y {k'}_y}_{20,12}\frac{1}{\varepsilon-\varepsilon_{1,2}-\Sigma_{1,2}(\varepsilon)+i0}V^{{k'}_y k_y}_{12,20}}\\
\Sigma_{1,2}(\varepsilon)&=&\disor{\sum_{{k'}_y}V^{k_y {k'}_y}_{12,12}\frac{1}{\varepsilon-\varepsilon_{1,2}-\Sigma_{1,2}(\varepsilon)+i0}V^{{k'}_y k_y}_{12,12}}\nonumber\\
&+&\disor{\sum_{{k'}_y}V^{k_y {k'}_y}_{12,20}\frac{1}{\varepsilon-\varepsilon_{2,0}-\Sigma_{2,0}(\varepsilon)+i0}V^{{k'}_y k_y}_{20,12}}
\end{eqnarray}

These coupled self consistent equations can be decoupled and transformed into quartic equations for either $\Sigma_{2,0}$ or $\Sigma_{1,2}$. They are cumbersome to solve except in special cases; for instance when one looks exactly the resonance field $B_r$ where the two unperturbed LL's cross i.e. $\varepsilon_{1,2}=\varepsilon_{2,0}$. If, in addition, we look at the center of the broadened LL's $\tilde{\varepsilon}=\varepsilon_{1,2}=\varepsilon_{2,0}$, we get :

\begin{eqnarray}
\Sigma_{2,0}\left(\tilde{\varepsilon}\right)&=&-\frac{\overline{V}_{20,20}^2}{\Sigma_{2,0}\left(\tilde{\varepsilon}\right)}-\frac{\overline{V}_{20,12}^2}{\Sigma_{1,2}\left(\tilde{\varepsilon}\right)}\\
\Sigma_{1,2}\left(\tilde{\varepsilon}\right)&=&-\frac{\overline{V}_{12,12}^2}{\Sigma_{1,2}\left(\tilde{\varepsilon}\right)}-\frac{\overline{V}_{12,20}^2}{\Sigma_{2,0}\left(\tilde{\varepsilon}\right)}\nonumber
\end{eqnarray}

where :
\begin{eqnarray}
\overline{V}_{12,20}^2&=\overline{V}_{20,12}^2=&\frac{x(1-x)\omega_0 \left(\Delta V\right)^2}{2\pi \lambda^2 L_z}\\
\overline{V}_{12,12}^2&=\overline{V}_{20,20}^2=&\frac{3}{2}\overline{V}_{12,20}^2\nonumber
\end{eqnarray}

The density of states at this particular energy $\tilde{\varepsilon}$ is equal to :

\begin{eqnarray}
\rho_t(\tilde{\varepsilon})=\rho_{2,0}(\tilde{\varepsilon})+\rho_{1,2}(\tilde{\varepsilon})&=\frac{L_xL_y}{2\pi \lambda^2}\frac{2}{\pi \sqrt{\overline{V}^2_{20,20}+\overline{V}^2_{20,12}}}
\end{eqnarray}

If there were no coupling between the $(2,0)$ and $(1,2)$ LL's and at energy $\tilde{\varepsilon}$ for either LL at  $B = B_r$ one would get : 
\begin{eqnarray}
\rho_{2,0}(\varepsilon=\tilde{\varepsilon})&=\frac{L_xL_y}{2\pi \lambda^2}\frac{1}{\pi \overline{V}_{20,20}}&=\rho_{1,2}(\varepsilon=\tilde{\varepsilon})
\end{eqnarray}

Hence, for these particular values of the energy $(\tilde{\varepsilon})$ and field ($B_r$) the ratio between the density of states  of the alloy - coupled LL's to that of either of the LL's neglecting interaction is :
\begin{eqnarray}
\frac{\rho_t}{\rho_{2,0}}=\frac{\rho_t}{\rho_{1,2}}&=2\sqrt{\frac{3}{5}}&\simeq 1.55
\end{eqnarray}

Similarly, it can be shown that at $B = B_r$, the DOS retains a semi - elliptical shape and extends over on each sides of $2 \sqrt{\overline{V}^2_{20,20}+\overline{V}^2_{20,12}}$.
We note that like in the isolated LL case the coupled LL width at resonance varies like $\sqrt{B_r}$.  When we numerically compute $\Delta$, the full width at $1/e$ maximum, we find a $\Delta (B = B_r = 30.9{\rm T}) \simeq  9.6 {\rm meV}$. Note that if there were only intra-$(2,0)$ contributions to the width of the Landau level, one would get $\Delta \simeq 7.5 {\rm meV}$ while if only inter $(2,0) - (1,2)$ contributions were retained this width would be $6.1 {\rm meV}$.  This demonstrates the efficiency of the alloy coupling between LL's belonging to different subbands and warns against an oversimplified addition of the scattering contributions to LL width (or collision frequencies).

\begin{figure}[!b]
\resizebox{.95\columnwidth}{!}{\includegraphics{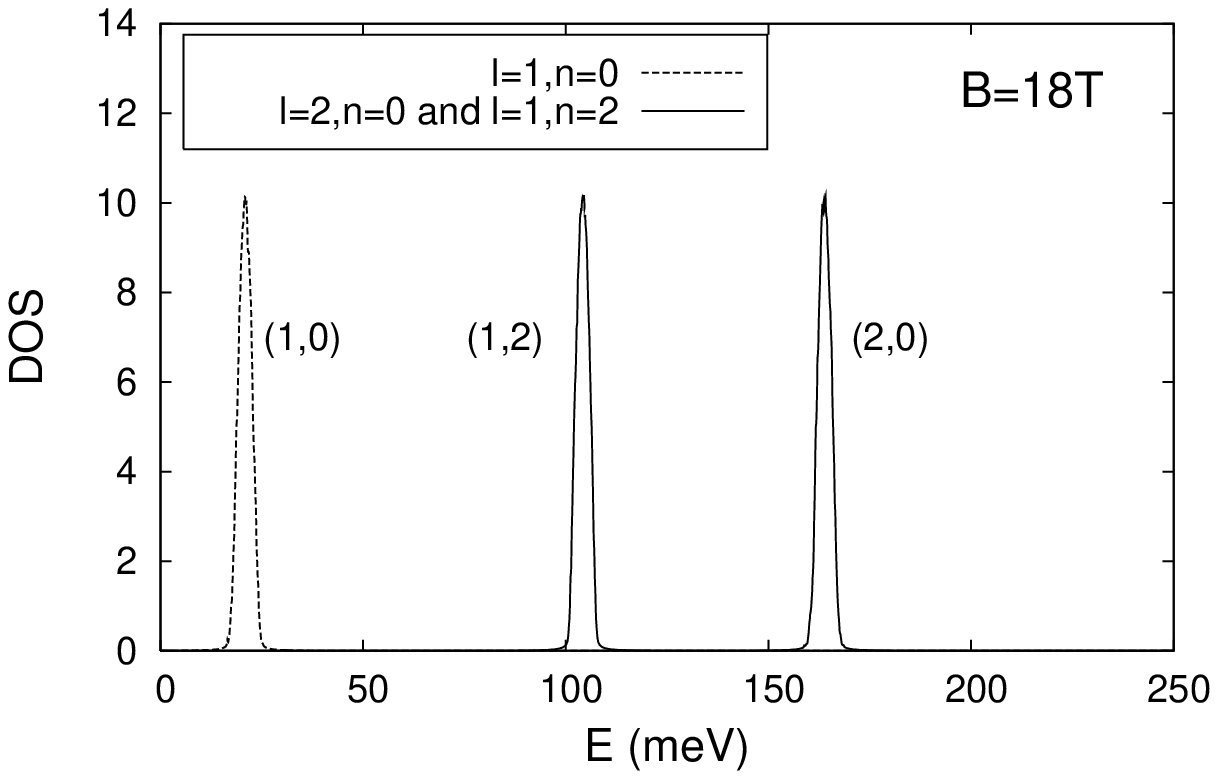}}
\resizebox{.95\columnwidth}{!}{\includegraphics{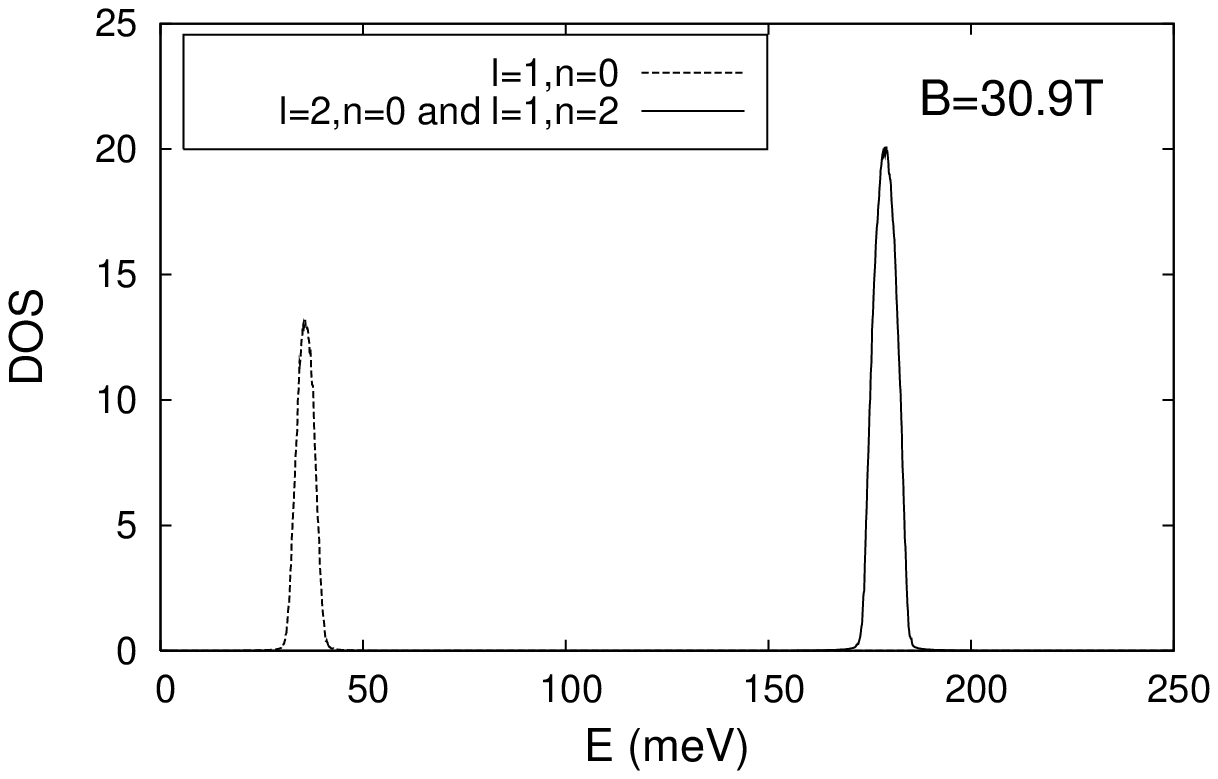}}
\resizebox{.95\columnwidth}{!}{\includegraphics{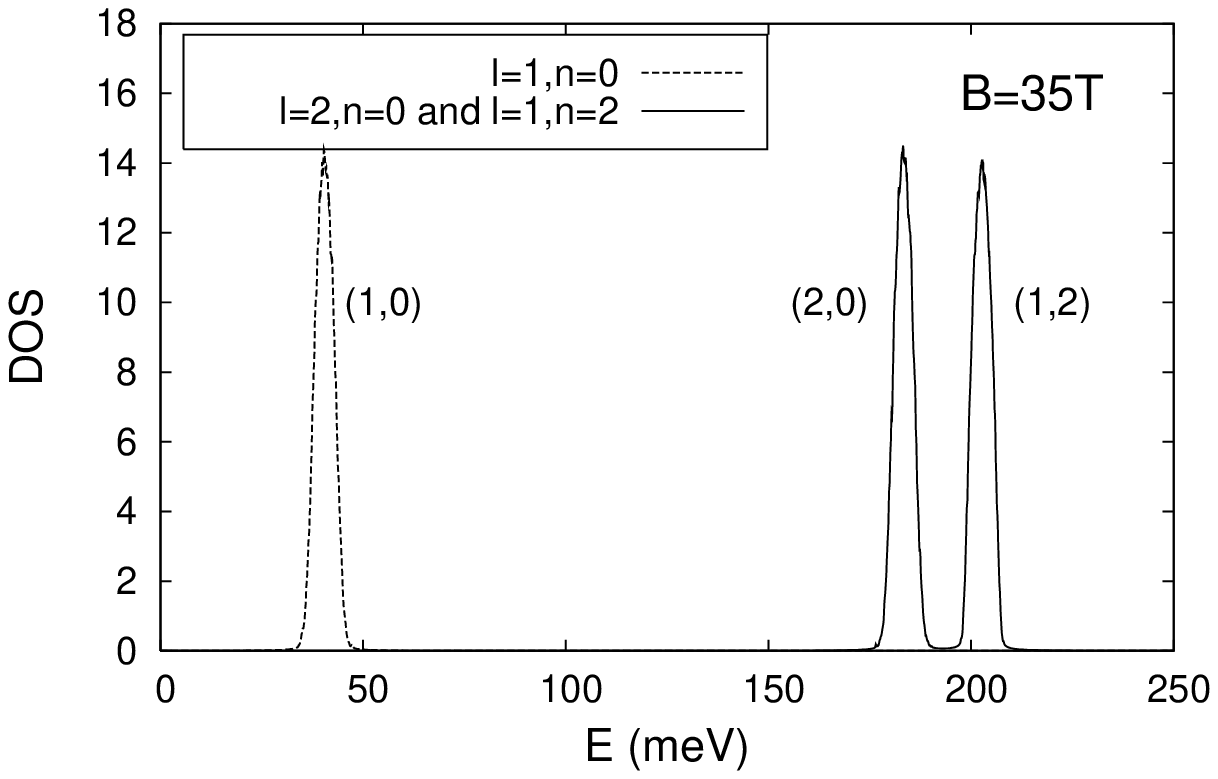}}
\caption{Density of states for different values of the magnetic field :  $B=18\; {\rm T}$ (top), $B=B_r=30.9\; {\rm T}$ (center) and $B=35\; {\rm T}$. (bottom). Solid lines correspond to the DOS considering both the $(2,0)$ and $(1,2)$ LL states, dashed lines are associated to the DOS with only $(1,0)$ LL states. For $B=18\; {\rm T}$ and $B=35\; {\rm T}$, we show the $(l,n)$ index of the states that mainly contribute to each DOS peak. Away from the resonance field $B_r$ (top and bottom figures), the shape of DOS depends neither on the LL nor on the subband.}
\label{dos}
\end{figure}

A numerical solution of the Schr\"odinger equation either inside a single LL or by taking into account the coupling between the two $(2,0)$ and $(1,2)$ LL's provides us with the eigenstates, eigenenergies and therefore DOS.  We show on the upper and lower panels of fig.\ref{dos} the numerically calculated DOS for non resonant fields (energy distances between LL's $\gg$ LL widths) for different LL's of different subbands. We see that they are almost identical, which confirms the reliability of the SCBA in the case of isolated LL's: not only is the DOS shape independent of the LL index but it is found to be also independent of the subband index. 

In the case of coupling and at resonance ($B = B_r$, middle panel in fig. \ref{dos}), we find that the LL's have acquired an extra width due to intersubband alloy scattering.  Indeed, as shown on fig.\ref{dos}, the height of the DOS peak is not twice that of a single peak.  The numerical ratio between the peak height of the coupled LL's (rhs peak) to that of a single LL (lhs peak) is $\simeq 1.53$.  This value is very close from the SCBA result ($2\sqrt{3/5}$) and again witnesses the reliability of that approach for short range scatterers, even including resonant couplings between LL's.

\section{Inter-subband magneto-absorption between alloy - broadened Landau levels}

The optical absorption probability from the $(1,0)$ LL states to the $(2,0)$ LL is closely related to the density of states and directly measurable. In the absence of disorder, the only allowed optical transition connects one $n = 0$ LL state of $E_1$ to the same LL state of $E_2$.  It is allowed for light propagating in the layer plane, i. e. in the $z$ polarization ($\meanvalbraket{\chi_1}{z}{\chi_2} \neq 0$). Since there is no non-parabolicity in the LL energies, the optical absorption probability spectrum $\alpha(\omega)$ of an ideal sample is a delta function :

\begin{eqnarray}
\alpha(\omega)&=&A\delta\left(\hbar\omega - \left(E_2-E_1\right)\right)\label{idealabsorption}
\end{eqnarray}

\begin{figure}[!b]
\resizebox{.95\columnwidth}{!}{\includegraphics{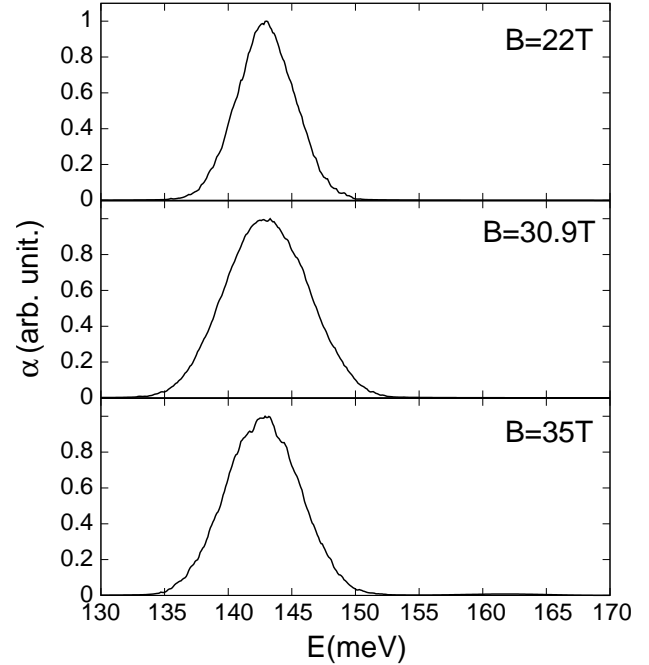}}
\caption{Absorption for different values of the magnetic field :  $B=22\; {\rm T}$ (top), $B=B_r=30.9\; {\rm T}$ (center) and $B=35\; {\rm T}$. (bottom)}
\label{absorption}
\end{figure}

We show in fig.\ref{absorption} the absorption lineshape calculated according to:

\begin{eqnarray}
\alpha(\omega)\propto\sum_{\nu,\nu'}& & \left|\left\langle\Psi_\nu |z|\Phi_{n=0,\nu'}\right\rangle\right|^2 \nonumber\\
& & \times \delta\left(\varepsilon_\nu - \varepsilon_{\nu'} - \hbar\omega\right)\label{disorderabsorption}
\end{eqnarray}

where $\nu$ labels the final eigenstates displayed in eq.\ref{DefPsiNu} while $\nu'$ labels the initial states $\left|\Phi_{n=0,\nu'}\right\rangle$ of the alloy-broadened $(1,0)$ LL.

\begin{figure}[!b]
\resizebox{.99\columnwidth}{!}{\includegraphics{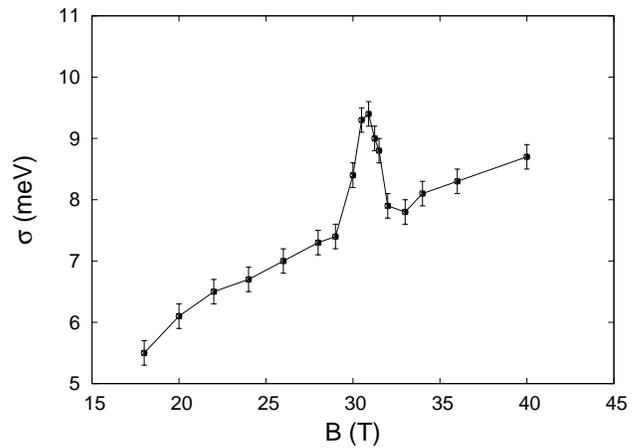}}
\caption{Absorption width as a function of the magnetic field. The solid line is a guide for the eyes.}
\label{broadening}
\end{figure}

In fig.\ref{absorption}, we see that the absorption lineshape in the vicinity of the $E_2-E_1$ energy resembles a broadened delta function.  Its full width at half maximum is shown in fig.\ref{broadening} versus $B$ where we see that increases roughly like $\sqrt{B}$ when the two interacting levels are sufficiently energy separated.  It is worth remarking that the DOS of each LL of the initial and final states has roughly the same width off resonance, in agreement with the SCBA predictions.  Instead, in the vicinity of the resonance the absorption width acquires an extra contribution (about $1.5 {\rm meV}$) due to the inter-level coupling induced by the alloy disorder.  The latter also activates optical transitions between the $(1,0)$ and the $(1,2)$ levels that are normally forbidden in ideal structures.  This cyclotron resonance harmonics occurs in the wrong polarization $(\epsilon_{\rm em} //z)$ and involves a $\Delta n = 2$ forbidden change in the LL index.  Fig. \ref{broadening} shows that this disorder - induced transition gains a substantial strength  in the vicinity of the $B_r$ field.

\section{Time dependent survival probability in the $(2,0$) Landau level}

One of the key issues in QCL physics is the quantitative understanding of the escape of the carriers from the upper state of the lasing transition. In a simplified rate equation approach, before lasing, the $(2,0)$ population $n_{20}$ is obtained by solving:

\begin{eqnarray}
\frac{dn_{20}}{dt}&=&g_{20}-n_{20}\left(\frac{1}{\tau_r}+\frac{1}{\tau_{nr}}\right)
\end{eqnarray}

The spontaneous emission power $W$ is found proportional to:

\begin{eqnarray}
W&\propto&\frac{1}{\tau_r}\left(\frac{1}{\tau_r}+\frac{1}{\tau_{nr}}\right)^{-1}
\end{eqnarray}

where $\tau_r$ et $\tau_{nr}$ are the population relaxation times of the upper level of the transition due to radiative and non radiative phenomena respectively. Around $B_r$, the radiative lifetime is not expected to vary much with an external magnetic field. On the other hand, the escape time due to non radiative processes involves elastic or inelastic scatterings from the $(2,0)$ LL to other levels (here $(1,n)$). Thus, the oscillatory output power of the QCL's subjected to a strong field is a natural consequence of the modulation of the electronic DOS by the field. Implicit in the previous analysis is the assumption that once a carrier has left $(2,0)$ it does not come back to $(2,0)$ because, once in $(1,2)$ (or whatever level) it will escape this level to a third one,..., much more quickly than returning to $(2,0)$.  In other words, it is assumed that the dynamics of the intra-subband relaxation $(1,2)$ to $(1,1)$ for instance is always faster that the dynamics of the inter-subband transitions.  This is often justified, in particular at zero field, because the density of states are continuous and the inter-subband matrix elements are consistently smaller than the intra-subband ones.  In QCL's, one faces instead a situation where the DOS presents gaps and it may turn out that an inter-subband channel opens while the intra-subband ones are non existent
In order to quantitatively analyze this aspect of the alloy - assisted escape processes, we reformulate the problem by projecting the 2 LL's hamiltonians as well as the alloy scattering potential on the 2 subspaces spanned by the 2 unperturbed LL's:

\begin{eqnarray}
H&=&H_0+V_{\rm alloy}\\
&=&{\cal P}_{2,0} \left(H_0+V_{\rm alloy}\right){\cal P}_{2,0} +{\cal P}_{1,2} \left(H_0+V_{\rm alloy}\right){\cal P}_{1,2}\nonumber\\
&&+{\cal P}_{2,0} V_{\rm alloy} {\cal P}_{1,2}+{\cal P}_{1,2} V_{\rm alloy} {\cal P}_{2,0}\nonumber
\end{eqnarray}

where ${\cal P}_{2,0}$ and ${\cal P}_{1,2}$ are the projectors on the unperturbed $(2,0)$ LL and $(1,2)$ LL respectively. The two first terms represent the intra-LL broadening due to alloy scattering while the last two terms induce transitions between the (broadened) LL's  We assume that at $t = 0$ the system has been placed in one of the eigenstates $\ket{l_0}$ of ${\cal P}_{2,0} \left(H_0+V_{\rm alloy}\right){\cal P}_{2,0}$.  As time flows, the system leaves this eigenstate.  While the intra-$(2,0)$ disorder is included in the basis of the eigenstates $\ket{l}$ of ${\cal P}_{2,0} \left(H_0+V_{\rm alloy}\right){\cal P}_{2,0}$, second (and higher) order couplings between these eigenstates exist due to the virtual jump(s) to the $(1,2$) subspace.  Thus, the survival probability in any particular eigenstate $\ket{l_0}$ very quickly drops to zero.  But this does not necessarily imply that the output of the QCL will be much affected if most of the transitions amount to keeping the state inside the ${\cal P}_{2,0} \left(H_0+V_{\rm alloy}\right){\cal P}_{2,0}$ subspace.  A good measure of the population relaxation time in the upper state of the lasing action is therefore provided by the calculation of the survival probability $P_s$ to find the system not in a given particular initial state $\ket{l_0}$ but in the whole ${\cal P}_{2,0} \left(H_0+V_{\rm alloy}\right){\cal P}_{2,0}$ subspace:

\begin{eqnarray}
P_s(t)&=\sum_l \mod{\braket{l}{\Psi(t)}}^2&=\sum_l \mod{\meanvalbraket{l}{e^{-iHt/\hbar}}{l_0}}^2\label{Psdef}
\end{eqnarray}

\begin{figure}[!b]
\resizebox{.95\columnwidth}{!}{\includegraphics{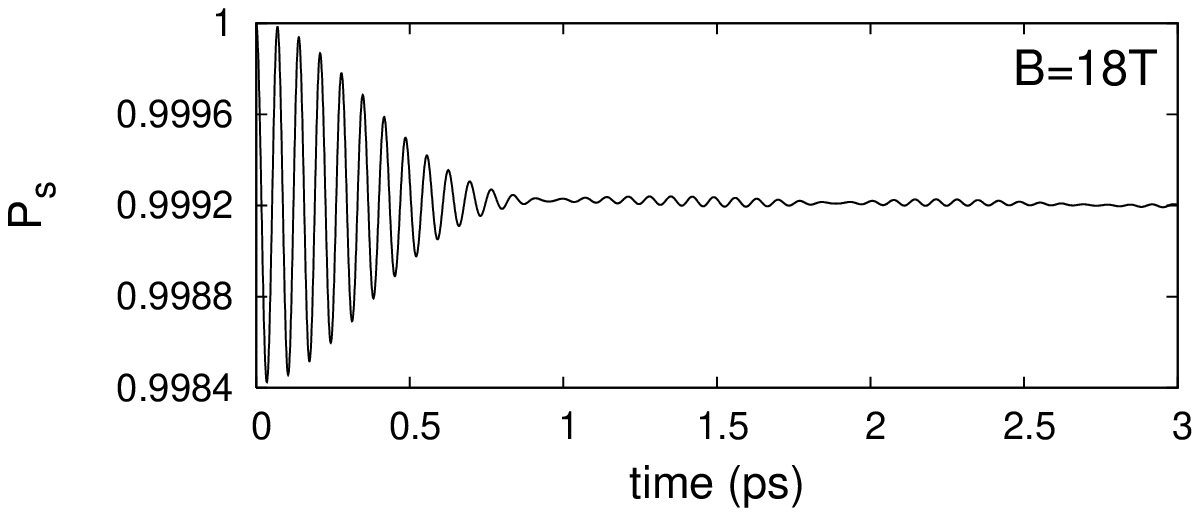}}
\resizebox{.95\columnwidth}{!}{\includegraphics{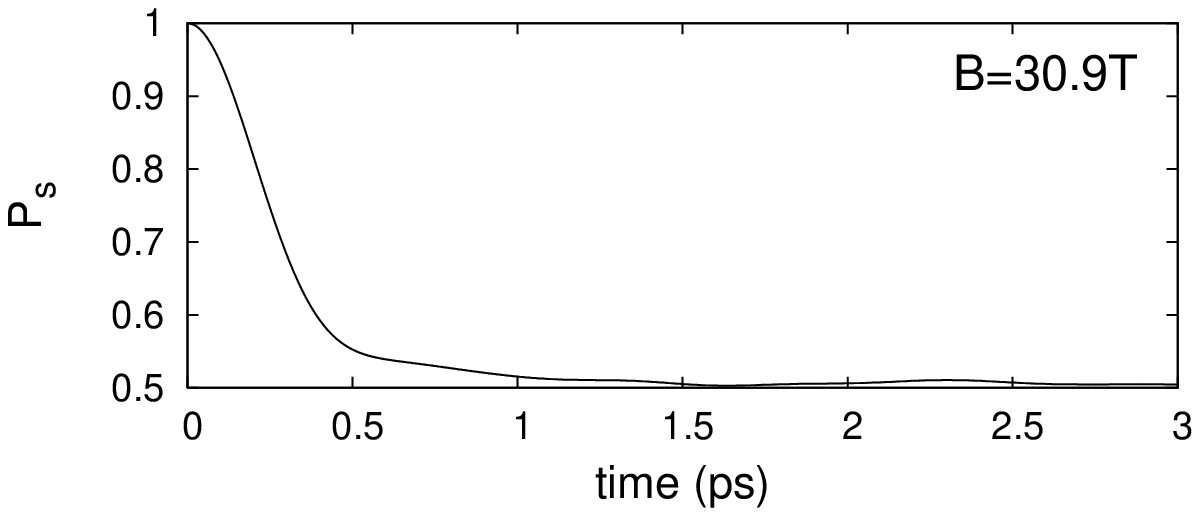}}
\resizebox{.95\columnwidth}{!}{\includegraphics{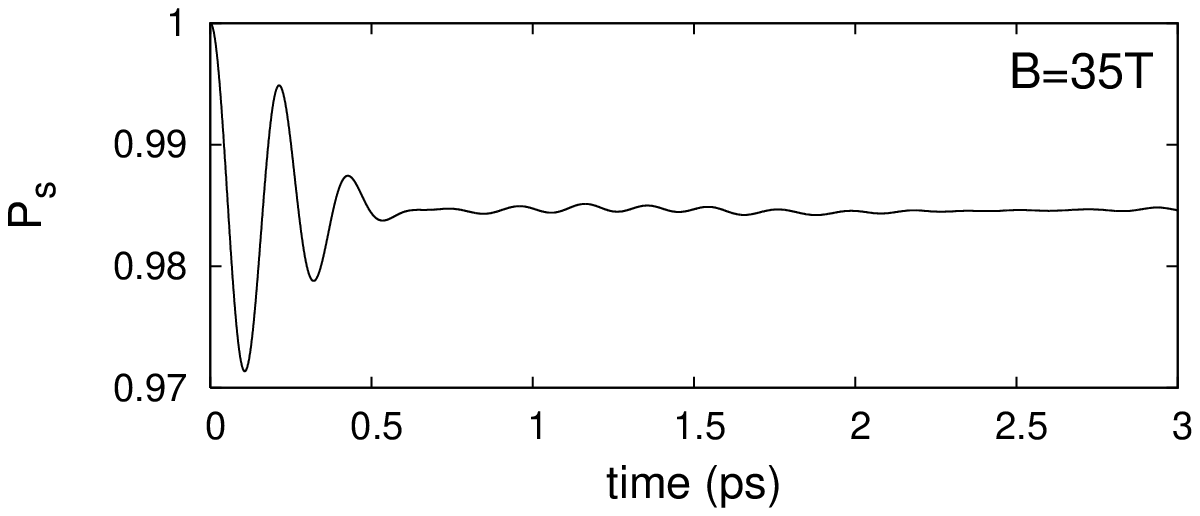}}
\caption{Survival probability $P_s$ of states for different values of the magnetic field :  $B=18\; {\rm T}$ (top), $B=B_r=30.9\; {\rm T}$ (center) and $B=35\; {\rm T}$. (bottom). The initial state is taken at the central level of the broadened $(2,0)$ LL's}
\label{escape}
\end{figure}

Fig.\ref{escape} shows the time dependence of $P_s(t)$ 
 when $\ket{l_0}$ is the central level of the broadened $(2,0)$ LL's for $B = 18{\rm T}$, $30.9{\rm T}$ and $35{\rm T}$.  These three fields correspond to off resonant situations (${\rm 18T}$ and ${\rm 35T}$) and the situation of exact resonance (${\rm 30.9T}$) respectively.  In the case of non resonant situations (fig. \ref{escape}, upper and lower panels) one finds fast beating between 1 (obtained at $t = 0$) and a value very close to 1: in practice there will not be any significant transfer between the two kinds of LL's and the laser will be bright if the only cause of depopulation of $(2,0)$ is the transfer assisted by alloy disorder to $(1,2)$. In fact, this behaviour is reminiscent of the time evolution of 1 level $\ket{0}$ coupled to $N$ degenerate levels $\ket{n}$. The survival probability in $\ket{0}$ is given by:

\begin{eqnarray}
P_0(t)&=&1-\frac{4\hbar^2\Omega^2}{\delta^2+4\hbar^2\Omega^2}\sin^2\left(\frac{t\sqrt{\delta^2+4\hbar^2\Omega^2}}{2\hbar}\right)\label{survivalprobability}
\end{eqnarray}

where $\delta$ is the detuning between $\ket{0}$ and the $N$ degenerate levels and $\hbar\Omega$ an effective coupling between $\ket{0}$ and these levels which is given by:

\begin{eqnarray}
\hbar\Omega&=&\sqrt{\sum_n \meanvalbraket{0}{V}{n}\meanvalbraket{n}{V}{0}}\label{effectivecouplng}
\end{eqnarray}

As applied to the QCL's, this simple modeling can work only if the detuning is large compared to the width of the LL's.  In fact, it correctly predicts that the product of the squared oscillation frequency by their amplitude should be proportional to $B$.  In addition, it shows that the period of the oscillations decreases with increasing detuning between the two LL's, in agreement with the numerical calculations.  However, the oscillations between 1 and $\delta^2/\left(\delta^2+4\hbar^2\Omega^2\right)$ in eq.\ref{survivalprobability} are everlasting while the numerical calculations exhibit a fast decay of the $P_s(t)$ envelope to a constant $1-\left(2\hbar^2\Omega^2\right)/\left(\delta^2+4\hbar^2\Omega^2\right)$.  This decay is clearly associated with the different Bohr frequencies of the problem: both the initial LL's have a finite width, thereby leading to many Bohr frequencies and thus to the replacement of the oscillatory sine function by an exponential decay. At resonance (fig.\ref{escape}, middle panel), the survival probability shows a very fast decay (subpicosecond) to a value $1/2$.  At large time, the survival probability in the initial subspace is just the fraction of states that belong to this subspace.  In our calculations there 74 states in each subset, thus the $1/2$ limiting value.

We have checked that the previous conclusions remain valid when the initial state is not the central state.  Actually, whatever the initial state, the calculated time constants (pseudo-period, characteristic transfer time at resonance) are independent of the initial state within a few \%.  This proves that our findings regarding a given initial state can be extended to the population of the initial Landau level irrespective of its precise nature (e.g. Boltzmann-like). In actual samples, there are of course other mechanisms that empty $(1,2)$ but do not refill the $(2,0)$ LL and finally, the survival probability in the initial subspace will of course goes to zero.  The key point is that the system leaves the initial subspace in a very short time.  Whether, in actual samples, this escape is irreversible or partly irreversible (which would occur if the escape from $E_1$, $n = 2$ is too slow) remains to be elucidated.  In the case of a quasi reversible escape, like fast elastic scatterings between LL's, we believe the population relaxation is better described by:

\begin{eqnarray}
P&=&\frac{g_{2,0}}{\tau_r}\left(1-\frac{2\hbar^2\Omega^2}{\delta^2+4\hbar^2\Omega^2}\right)
\end{eqnarray}

where $\delta=E_{2,0}-E_{1,2}$ and $\hbar\Omega$ is an effective coupling constant between the $(2,0)$ and $(1,2)$ LL's and more generally between $(2,0)$ and $(1,n)$.  A minimum is clearly reached at resonance, but this minimum is non zero.  To suppress the QCL emission, it is necessary to include other losses. A preliminary version of this work was presented at the ICPS 28 Vienna (2006).

\begin{acknowledgments}
We thank Angela Vasanelli and Carlo Sirtori for fruitful discussions. The LPA-ENS is `laboratoire associ\'e au CNRS UMR 8551 et aux Universit\'es Pierre et Marie Curie (Paris 6) et Denis Diderot (Paris 7)'. 

\end{acknowledgments}

\end{document}